\documentclass[dvipsnames]{article} %
\usepackage{colm2024_conference}
\pdfoutput=1
\usepackage{natbib}
\setcitestyle{authoryear,round,citesep={;},aysep={,},yysep={;}}
\let\cite\citep
\usepackage{booktabs}
\usepackage{graphicx}
\usepackage{enumitem}
\usepackage{wrapfig}
\usepackage{algorithm}
\usepackage{algpseudocode}
\usepackage{microtype}
\usepackage{amsmath}
\usepackage{amsthm}
\usepackage{colortbl}
\usepackage[utf8]{inputenc}
\definecolor{lightgray}{rgb}{0.9,0.9,0.9}
\usepackage{caption}
\usepackage{subcaption}
\usepackage{setspace}
\usepackage{url}
\usepackage{multirow}
\usepackage{colortbl}
\usepackage{tabularx}
\usepackage{blindtext}
\usepackage{pgfplots}
\pgfplotsset{compat=1.18} 
\usepackage{tikz}
\usetikzlibrary{er,positioning,bayesnet}
\usepackage{makecell}
\usepackage{tipa}
\usepackage{siunitx}
\usepackage{nicefrac}
\usepackage{tocloft}
\usepackage{listings}
\usepackage{cleveref}
\usepackage{placeins}
\usepackage{rotating}
\usepackage{silence}
\usepackage{tablefootnote}
\usepackage{threeparttable}
\usepackage[raster,skins]{tcolorbox} %
\usepackage{xltabular}
\usepackage{adjustbox}
\usepackage{xurl}
\usepackage[normalem]{ulem}
\useunder{\uline}{\ul}{}
\usepackage{newtxtext,newtxmath}
\usepackage{CJKutf8}
\usepackage{pifont}
\usepackage{xcolor}

\usepackage{amsmath,amsfonts,bm}

\def\eqref#1{equation~\ref{#1}}

\def\1{\bm{1}}

\DeclareMathAlphabet{\mathsfit}{\encodingdefault}{\sfdefault}{m}{sl}
\SetMathAlphabet{\mathsfit}{bold}{\encodingdefault}{\sfdefault}{bx}{n}

\newcommand*\justify{%
  \fontdimen2\font=0.4em%
  \fontdimen3\font=0.2em%
  \fontdimen4\font=0.1em%
  \fontdimen7\font=0.1em%
  \hyphenchar\font=`\-%
}

\renewcommand{\texttt}[1]{%
  \begingroup
  \ttfamily
  \begingroup\lccode`~=`/\lowercase{\endgroup\def~}{/\discretionary{}{}{}}%
  \begingroup\lccode`~=`[\lowercase{\endgroup\def~}{[\discretionary{}{}{}}%
  \begingroup\lccode`~=`.\lowercase{\endgroup\def~}{.\discretionary{}{}{}}%
  \catcode`/=\active\catcode`[=\active\catcode`.=\active
  \justify\scantokens{#1\noexpand}%
  \endgroup
}

\usepackage[table]{xcolor} % 必须要有这个包
\definecolor{pyellow}{HTML}{FFFACD}
\definecolor{mygreen}{HTML}{006400}

\title{\centering MoE Adapter for Large Audio Language Models: Sparsity, Disentanglement, and Gradient-Conflict-Free}

\author{
    \centering
    \small
    \textbf{Yishu Lei$^{*1}$}\quad
    \textbf{Shuwei He$^{*1, 2}$}\quad
    \textbf{Jing Hu$^{1, 3}$}\quad
    \textbf{Dan Zhang$^{1}$}\quad
    \textbf{Xianlong Luo$^{1}$}\quad
    \textbf{Danxiang Zhu$^{1}$} \\
    \textbf{Shikun Feng$^{\dagger1}$}\quad
    \textbf{Rui Liu$^{2}$}\quad
    \textbf{Jingzhou He$^{1}$}\quad
    \textbf{Yu Sun$^{1}$}\quad
    \textbf{Hua Wu$^{1}$}\quad
    \textbf{Haifeng Wang$^{1}$}
    \\[0.8em]
    \small
    $^{1}$ERNIE Team, Baidu \\
    $^{2}$College of Computer Science, Inner Mongolia University \\
    $^{3}$Tsinghua Shenzhen International Graduate School, Tsinghua University
    \\[0.8em]
    \small
    \texttt{\{leiyishu, heshuwei\}@baidu.com, cminuser@gmail.com,} \\
    \texttt{\{zhangdan20, luoxianlong, zhudanxiang\}@baidu.com,} \\
    \texttt{fengshikun01@baidu.com, imucslr@imu.edu.cn,} \\
    \texttt{\{hejingzhou, sunyu02, wu\_hua, wanghaifeng\}@baidu.com} \\
    \vspace{0.5em}
    {\footnotesize $^{*}$Equal contribution \quad $^{\dagger}$Corresponding author}
}

\makeatletter\def\@abstract{
Extending the input modality of Large Language Models~(LLMs) to the audio domain is essential for achieving comprehensive multimodal perception.
However, it is well-known that acoustic information is intrinsically \textit{heterogeneous}, entangling attributes such as speech, music, and environmental context.
Existing research is limited to a dense, parameter-shared adapter to model these diverse patterns, which induces \textit{gradient conflict} during optimization, as parameter updates required for distinct attributes contradict each other.
To address this limitation, we introduce the \textit{\textbf{MoE-Adapter}}, a sparse Mixture-of-Experts~(MoE) architecture designed to decouple acoustic information.
Specifically, it employs a dynamic gating mechanism that routes audio tokens to specialized experts capturing complementary feature subspaces while retaining shared experts for global context, thereby mitigating gradient conflicts and enabling fine-grained feature learning.
Comprehensive experiments show that the MoE-Adapter achieves superior performance on both audio semantic and paralinguistic tasks, consistently outperforming dense linear baselines with comparable computational costs. 
Furthermore, we will release the related code and models to facilitate future research.}\makeatother

\begin{document}
\maketitle
% {\let\thefootnote\relax\footnotetext{$^*$Equal contribution, $^\dag$Corresponding author}}
% \newpage

% \clearpage
\pagestyle{firstpage}  %
% \tableofcontents
% \clearpage
\pagestyle{normalpage}

\section{Introduction}

\begin{figure}[htbp]
    \centering
    \includegraphics[width=0.7\linewidth]{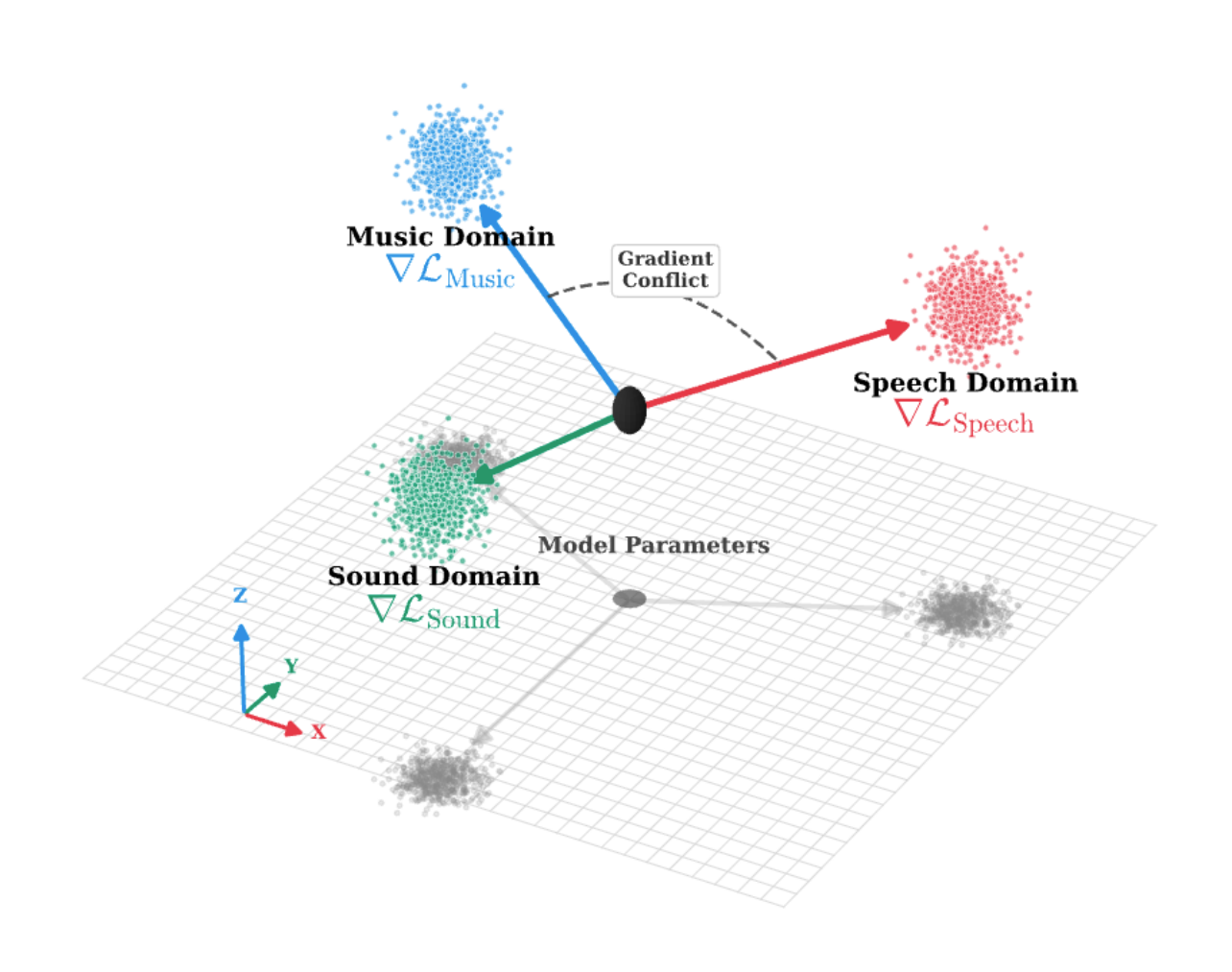}
    \caption{\textbf{Visualization of Gradient Conflict.} The naturally disjoint manifolds of Speech, Music, and Sound induce divergent gradient update directions ($\nabla \mathcal{L}$), creating an optimization conflict that pulls the shared adapter in contradictory ways.}
    \label{fig:heterogeneity}
\end{figure}

Large Language Models~(LLMs) have revolutionized cognitive intelligence through their exceptional text-processing capabilities~\cite{gpt4, deepseek-r1, qwen3}. 
However, limiting these models to the textual modality constrains their ability to perceive and interact with the dynamic physical world. 
As audio serves as a primary channel for both human communication and environmental perception, extending LLMs to the audio domain is a critical research frontier~\cite{audiopalm, wavchat, ltu}.
The prevailing paradigm involves inserting an adapter to project acoustic features into the textual semantic space.
However, recent approaches predominantly rely on a dense, parameter-shared adapter for this purpose~\cite{salmonn, qwen2-audio, mini-omni, longcat-flash-omni, qwen3-omni}. 
For instance, Kimi-Audio~\cite{kimi-audio} utilizes a fixed MLP-based projector, while Audio Flamingo 3~\cite{audio-flamingo3} employs temporal compression layers with globally shared parameters. 
While straightforward, these designs implicitly assume that a single unified projection can uniformly capture the intrinsic heterogeneity of complex audio signals.

However, as illustrated in Figure~\ref{fig:heterogeneity}, this assumption conflicts with the intrinsic topology of the acoustic modality: rather than conforming to a uniform distribution, human speech, music, and environmental sounds occupy distinct manifolds with divergent statistical properties~\cite{audioset, superb}. Constraining a static, parameter-shared adapter to map these heterogeneous inputs into a unified space precipitates significant gradient conflicts~\cite{gradnorm, pcgrad}. Specifically, parameter updates required for semantic speech often contradict those for paralinguistic cues, engendering destructive interference within the shared weights~\cite{conflict-averse}. We empirically substantiate these conflicts in Section~\ref{sec:analysis}. Consequently, a monolithic adapter is ill-suited for simultaneous alignment and disentanglement. This raises a fundamental question: 
\textit{How can an architecture be designed to explicitly accommodate the heterogeneous structure of audio data rather than suppressing it?}

Mixture-of-Experts~(MoE) architectures offer a principled inductive bias for managing such intrinsic distributional heterogeneity. 
By conditionally activating sparse subsets of experts, MoE enables the decomposition of complex data manifolds into manageable subspaces, allowing distinct experts to specialize in specific acoustic attributes~\cite{shazeer2017outrageously, switch-transformer}. 
This specialization effectively orthogonalizes optimization directions, mitigating gradient conflicts arising from entangled features.
While the efficacy of this paradigm has been rigorously validated in vision and multi-modal domains~\cite{vmoe, limoe, zhang2025clip, li2025uni, su2025hamobe}, its potential remains largely untapped in audio-text alignment. 
To date, prevalent Large Audio Language Models~(LALMs) persist in their reliance on static, monolithic adapters, leaving the challenge of acoustic heterogeneity unaddressed.

To bridge this gap, we introduce the \textit{\textbf{MoE-Adapter}}, a sparse and dynamic architecture explicitly designed to resolve optimization bottlenecks induced by acoustic heterogeneity. 
Departing from the monolithic paradigm that subjects all parameters to uniform global updates, our approach leverages a learnable gating mechanism to dynamically route input segments to specialized experts based on their intrinsic acoustic attributes. 
This mechanism orchestrates a decoupled optimization process: gradient updates from conflicting signal types are isolated to specific experts, significantly reducing interference while retaining shared experts for common feature extraction. 
Consequently, the MoE-Adapter achieves effective disentanglement of diverse acoustic factors and alignment with the textual space, all while maintaining inference-time computational costs comparable to dense baselines.

In summary, our main contributions are three-fold:
\begin{itemize}
    \item \textbf{Architectural Innovation:} We propose the MoE Adapter, a sparse, dynamically routed projection layer that seamlessly replaces static adapters, explicitly targeting gradient conflicts via expert specialization.
    \item \textbf{Empirical Insight:} We substantiate the existence of severe gradient interference across semantic, paralinguistic, and environmental factors in parameter-shared adapters, and demonstrate how our method effectively alleviates these optimization bottlenecks.
    \item \textbf{Performance \& Efficiency:} 
    We validate the proposed architecture on a 1.7B-parameter LLM backbone. 
    The MoE-Adapter consistently outperforms dense baselines in both semantic understanding and paralinguistic tasks, 
    demonstrating a more favorable performance--efficiency trade-off under a fixed model capacity.
    
\end{itemize}

\section{Related Works}

\subsection{Large Audio Language Models}
LALMs have evolved from cascaded pipelines to unified end-to-end frameworks~\cite{wavchat}.
Early cascaded models~\cite{speechgpt, audiogpt} relied on Automatic Speech Recognition~(ASR) transcripts, suffering from error propagation and the loss of non-verbal cues.
In contrast, end-to-end models~\cite{audiopalm, ltu} bridge the modality gap by mapping acoustic features to the textual space via learnable adapters.
These adapters typically fall into two categories: Q-Former-based architectures, exemplified by SALMONN~\cite{salmonn}, and Linear Projectors, as adopted by Qwen2-Audio~\cite{qwen2-audio}, GLM-4-Voice~\cite{glm4voice}, LLaMA-Omni~\cite{llama-omni}, Step-Audio2~\cite{step-audio2} and Kimi-Audio~\cite{kimi-audio}.
While simpler and more efficient, current state-of-the-art models largely favor the latter, utilizing a dense, global projection layer.
However, this prevailing design implicitly assumes a uniform distribution across diverse audio types.
As a result, it struggles to resolve the distributional shifts between semantic speech and acoustic events, leading to gradient conflicts~\cite{pcgrad, modality-conflict, gapo}.
In this work, we challenge this monolithic convention by introducing a sparse MoE mechanism.

\subsection{Mixture-of-Experts Architectures}

The MoE paradigm has evolved from decomposing complex problems~\cite{jacobs1991adaptive} into a standard for efficient training that handles data heterogeneity and mitigates gradient conflicts via sparse gating~\cite{shazeer2017outrageously}. 
Recent studies in multimodal domains~\cite{chen2023octavius, li2025uni, zhang2025clip} validate this capability, demonstrating that routing conflicting tasks to orthogonal experts effectively resolves optimization bottlenecks where different modalities require contradictory parameter updates~\cite{huang2025moe, xin2025i2moe}.In the audio domain, the adoption of MoE is similarly expanding. 
Recent works have successfully adapted this paradigm for specific applications, exemplified by speech and music generation~\cite{liu2025unimoe} and robust feature selection in healthcare~\cite{shao2025motas}.
However, despite these advancements, the potential of MoE for general-purpose audio--text alignment remains largely untapped.
Prevalent LALMs persist in relying on static, parameter-shared adapters—a monolithic design that forces a single set of parameters to model intrinsically diverse acoustic attributes.
We address this limitation by introducing the MoE-Adapter, leveraging dynamic expert routing to explicitly disentangle conflicting acoustic patterns during the alignment phase.

\section{Method}
\label{sec:method}

\begin{figure*}[t!]
  \centering
  \includegraphics[width=\linewidth]{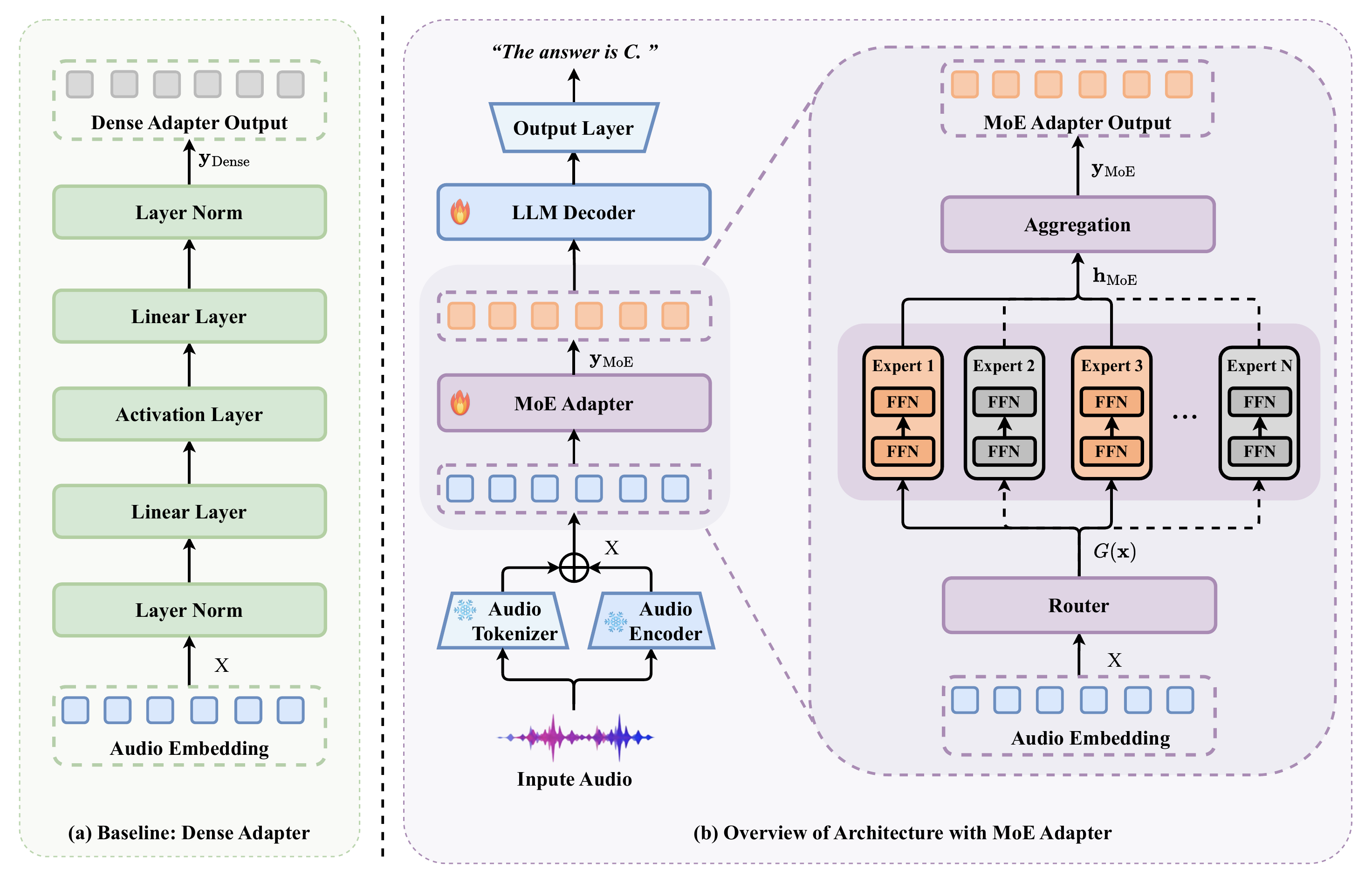}
    \caption{\textbf{The overall architecture of the MoE Adapter compared with the Dense Adapter baseline.} Unlike the monolithic, parameter-shared design in~(a), the MoE-Adapter in~(b) employs a sparse gating mechanism to dynamically route audio tokens to specialized experts. This architecture facilitates fine-grained feature learning and effectively mitigates gradient conflicts by disentangling heterogeneous acoustic attributes during the projection phase.}
  \label{fig:architecture}
\end{figure*}

\subsection{Overall Framework}
\label{ssec:framework}

As depicted on the left side of Figure~\ref{fig:architecture}~(b), our model leverages the dual-stream architecture of Kimi-Audio~\cite{kimi-audio}. 
The audio frontend employs two pathways: a frozen tokenizer extracts \textit{discrete semantic tokens}, while a speech encoder captures \textit{continuous acoustic features}. 
These heterogeneous representations are subsequently aligned via a projection layer and fused through element-wise summation. 
Instead of using a standard dense adapter, we employ a MoE-Adapter to dynamically map the fused features into the LLM's embedding space.
Finally, the adapted audio embeddings are concatenated with text token embeddings and serve as input to the backbone for standard autoregressive next-token prediction~(NTP).

\subsection{Sparse MoE Adapter}
\label{ssec:moe_layer}

\paragraph{Preliminary: Dense Adapter.}
As shown in Figure~\ref{fig:architecture}~(a), the LALMs typically employ a monolithic Feed-Forward Network~(FFN) as the modality projector. 
Formally, given an input audio token $\mathbf{x} \in \mathbb{R}^d$, the projected embedding $\mathbf{y} \in \mathbb{R}^D$ is formulated as:
\begin{equation}
    \mathbf{y} = \mathcal{N}\Big(\mathbf{W}_{d2} \cdot \sigma\big(\mathbf{W}_{d1} \cdot \mathcal{N}(\mathbf{x})\big)\Big),
    \label{eq:dense_adapter}
\end{equation}
where $\mathcal{N}(\cdot)$ denotes Layer Normalization, $\sigma(\cdot)$ represents the SiLU activation function, and $\mathbf{W}_{d1}, \mathbf{W}_{d2}$ are the learnable weight matrices specific to the dense projector.
This output $\mathbf{y}$ is subsequently utilized to substitute audio placeholder tokens within the LLM input sequence.
However, this monolithic design imposes a rigid inductive bias: a single set of parameters is forced to accommodate heterogeneous acoustic modalities, inevitably inducing optimization interference.

\paragraph{Sparse Adapter with MoE Mechanisms.}
Figure~\ref{fig:architecture}~(b) illustrates the overall architecture of the MoE Adapter.
To facilitate fine-grained feature specialization, we replace the monolithic architecture with a bank of $N$ experts, denoted as $\{E_i\}_{i=1}^N$. 
Each expert $E_i$ is parameterized as a lightweight FFN utilizing the SiLU activation $\phi(\cdot)$:
\begin{equation}
    E_i(\mathbf{x}) = \mathbf{W}_{e2}^{(i)} \cdot \phi\big(\mathbf{W}_{e1}^{(i)} \cdot \mathcal{N}(\mathbf{x})\big),
    \label{eq:expert_fn}
\end{equation}
where $\mathbf{W}_{e1}^{(i)}$ and $\mathbf{W}_{e2}^{(i)}$ denote the unique weight matrices for the $i$-th expert.
To orchestrate expert selection, a learnable router $G(\cdot)$ computes sparse gating probabilities based on the logits $\mathbf{s} = \mathbf{x}\mathbf{W}_g$, where $\mathbf{W}_g$ is the gating matrix:
\begin{equation}
    G(\mathbf{x}) = \operatorname{softmax}\big(\mathcal{T}_k(\mathbf{s})\big).
    \label{eq:gating}
\end{equation}
Here, $\mathcal{T}_k(\cdot)$ represents the Top-$k$ operator, which retains the $k$ largest values while masking the remainder to $-\infty$. 
The expert outputs are synthesized via weighted aggregation to yield the intermediate representation $\mathbf{h}_{\text{MoE}}$:
\begin{equation}
    \mathbf{h}_{\text{MoE}} = \sum_{i \in \mathcal{I}} G(\mathbf{x})_i \cdot E_i(\mathbf{x}),
    \label{eq:moe_aggregation}
\end{equation}
where $\mathcal{I}$ denotes the set of indices corresponding to the active experts selected by the router.

\paragraph{Output Projection and Aggregation.}
To align the dimension of the aggregated features with the LLM's embedding space, $\mathbf{h}_{\text{MoE}}$ is processed by a final linear projection layer followed by layer-normalization:
\begin{equation}
    \mathbf{y}_{\text{MoE}} = \mathcal{N}\big(\mathbf{W}_P \cdot \mathbf{h}_{\text{MoE}}\big),
    \label{eq:projection}
\end{equation}
where $\mathbf{W}_P$ is the learnable projection matrix. 
This adapted output $\mathbf{y}_{\text{MoE}}$ is used to replace the pre-defined audio placeholders in the input sequence, thereby conditioning the next-token prediction task on the optimized, disentangled acoustic context.

\subsection{Training Objectives}

The entire model is optimized end-to-end using a joint objective:
\begin{equation}
\mathcal{L} = \mathcal{L}_{\text{NTP}} + \lambda \mathcal{L}_{\text{aux}}.
\end{equation}

\paragraph{Next-Token Prediction Loss.}
The primary training objective is next-token prediction conditioned on the audio context:
\begin{equation}
\mathcal{L}_{\text{NTP}} = - \sum_{t=1}^{T} \log P(y_t \mid y_{<t}, \mathbf{X}; \theta),
\end{equation}
where $\theta$ denotes the model parameters and $\mathbf{X}$ refers to the fused audio representation before the adapter, whose adapted form is used to condition the LLM.

\paragraph{Auxiliary Load-Balancing Loss.}
To mitigate expert collapse, we adopt a standard load-balancing loss~\cite{gshard} over the routed experts $\mathcal{E}_R$.
Let $B$ denote the number of routed tokens in a batch.
For each token $\mathbf{x}_b$, the router outputs a routing probability
$p_{b,e} = G(\mathbf{x}_b)_e$ for expert $e \in \mathcal{E}_R$.
We define the \emph{expert importance} as the average routing probability,
\begin{equation}
\bar{P}_e = \frac{1}{B} \sum_{b=1}^{B} p_{b,e},
\end{equation}
and the \emph{expert load} as the fraction of tokens routed to expert $e$,
\begin{equation}
\bar{f}_e = \frac{1}{B} \sum_{b=1}^{B} r_{b,e},
\end{equation}
where $r_{b,e}=1$ if expert $e$ is selected for token $\mathbf{x}_b$ by Top-$k$ routing, and $0$ otherwise.
The auxiliary loss is then computed as:
\begin{equation}
\mathcal{L}_{\text{aux}} = |\mathcal{E}_R| \sum_{e \in \mathcal{E}_R} \bar{P}_e \cdot \bar{f}_e.
\end{equation}

\section{Experiments}

\subsection{Experimental Setups}

\paragraph{Model Architecture \& Training Settings.}
The LLM backbone utilizes Qwen3-1.7B~\footnote{\url{https://huggingface.co/Qwen/Qwen3-1.7B}}. 
The audio frontend employs the Whisper-VQ tokenizer~\cite{glm4voice} and the Whisper Encoder~\cite{whisper}.
Training is conducted on a high-quality 40B-token corpus using AdamW~($\beta_1=0.9, \beta_2=0.95$) and a Warmup-Stable-Decay scheduler~\cite{wen2024understanding}~(peak LR $1\times 10^{-5}$, 20 warmup steps). 
For fair comparison~\footnote{Additional training hyperparameters and implementation details are provided in Appendix.}, we constrain the total parameter budget of both the dense adapter and the MoE-Adapter to 94.4M parameters. 
Notably, due to sparse activation, the MoE-Adapter activates only approximately 70.8M parameters during inference, corresponding to about 75\% of the baseline. 
In addition, we fix random seeds across all experiments to control stochasticity and ensure experimental fairness.

\paragraph{Evaluation.}
We conduct few-shot evaluations on widely adopted benchmarks.
Specifically, we leverage MMAU~\cite{mmau}—a large-scale benchmark covering speech, sounds, and music—for \textit{perception-oriented audio understanding}, focusing on acoustic and paralinguistic reasoning.
For world knowledge reasoning, we utilize the MMSU and OpenBookQA~(OBQA) subsets of VoiceBench~\cite{voicebench}, which are audio adaptations of the text-only MMLU-Pro~\cite{mmlu-pro} and require \textit{higher-level semantic reasoning} beyond low-level perceptual cues.
All evaluations are performed using greedy decoding to ensure deterministic and comparable results.

\subsection{Main Results}

Table~\ref{tab:main_comprehensive} presents the performance comparison between our method and the baseline across all benchmarks. Here, we highlight how the MoE-Adapter drives consistent improvements across three key dimensions: knowledge reasoning, paralinguistic understanding, and cross-modal alignment.

\definecolor{pyellow}{HTML}{FFFACD}
\definecolor{mygreen}{HTML}{006400}

\begin{table}[htbp]
\centering
\small
\setlength{\tabcolsep}{8pt}

\caption{\textbf{Main results on audio understanding and cross-modal alignment.} 
Compared to a dense FFN baseline under a controlled budget, MoE Adapter consistently improves audio accuracy and reduces the modality gap (the performance disparity between audio and text inputs), demonstrating superior representational capacity.}
\label{tab:main_comprehensive}

\resizebox{\linewidth}{!}{%
\begin{tabular}{l c ccc cc}
\toprule
\multirow{3}{*}{\textbf{Benchmark}} 
& \textbf{Text Accuracy}~($\uparrow$)
& \multicolumn{3}{c}{\textbf{Audio Accuracy} ($\uparrow$)} 
& \multicolumn{2}{c}{\textbf{Modality Gap} ($\uparrow$)} \\
\cmidrule(lr){2-2} \cmidrule(lr){3-5} \cmidrule(lr){6-7}
& Backbone LLM 
& Baseline & \textbf{Ours} & \textit{Improv.} 
& Baseline & \textbf{Ours} \\

\midrule
\rowcolor{pyellow} 
\multicolumn{7}{l}{\textit{\textbf{Task: Audio Knowledge Reasoning}}} \\ 
\midrule

MMSU~\cite{voicebench} & 52.86 & 35.03 & \textbf{38.19} & \textbf{\textcolor{mygreen}{+3.16}} & -17.83 & \textbf{-14.67} \\
OBQA~\cite{voicebench} & 68.35 & 50.10 & \textbf{53.85} & \textbf{\textcolor{mygreen}{+3.75}} & -18.25 & \textbf{-14.50} \\

\midrule
\rowcolor{pyellow}
\multicolumn{7}{l}{\textit{\textbf{Task: Audio Paralinguistic Tasks}}} \\
\midrule

MMAU~\cite{mmau}       & -     & 59.79 & \textbf{61.50} & \textbf{\textcolor{mygreen}{+1.71}} & -      & - \\

\bottomrule
\end{tabular}
}
\end{table}

\paragraph{Knowledge Reasoning Gains.}
Our method consistently outperforms the baseline across audio knowledge benchmarks. Specifically, on MMSU and OBQA, we achieve gains of \textbf{3.16\%} ($35.03\% \rightarrow 38.19\%$) and \textbf{3.75\%} ($50.10\% \rightarrow 53.85\%$), respectively. These results suggest that expert specialization enables the model to effectively capture knowledge-centric semantic information, thereby enhancing both commonsense reasoning and cross-domain knowledge integration.

\paragraph{Robust Paralinguistic Understanding.}
Beyond knowledge tasks, our method demonstrates robust performance on paralinguistic understanding, achieving a \textbf{1.71\%} ($59.79\% \rightarrow 61.50\%$) improvement on the MMAU benchmark. This indicates that the MoE mechanism effectively learns diverse acoustic cues, enabling higher-level reasoning in complex scenarios and demonstrating robustness across heterogeneous audio domains.

\paragraph{Reduced Modality Gap.}
We further quantify audio--text consistency using the Modality Gap, where our method shows consistent reductions. For instance, on MMSU, the gap narrows significantly from $-17.83$ to $-14.67$ (an improvement of \textbf{3.16}). These gains confirm that the dynamic routing mechanism maps heterogeneous acoustic representations into subspaces that are more compatible with the LLM of text embedding space, yielding superior alignment compared to static dense adapters.

\subsection{Ablation Studies}
\label{sec:ablation}

In this section, we conduct ablation studies to examine the effectiveness of key architectural and routing design choices in MoE-Adapter.
We analyze how expert configuration, routing sparsity, and training objectives influence audio understanding and reasoning performance across different benchmarks.

\subsubsection{Expert Configuration and Sparsity}
We analyze how expert configuration and routing sparsity affect audio understanding and reasoning under the same training token budget. Results on MMAU, MMSU, and OBQA are summarized in Table~\ref{tab:ablation_experts_40b}.

\begin{table*}[htbp] 
    \centering
    \small
    \setlength{\tabcolsep}{15pt} 
    
    \caption{\textbf{Ablation on expert configurations.} We analyze the impact of expert count, routing sparsity, and dimension capacity. The default setting ``8 choose 4'' (highlighted) achieves the best overall balance.}
    \label{tab:ablation_experts_40b}
    
    \begin{tabular}{lc ccc}
    \toprule
    \multirow{2}{*}{\textbf{Configuration}} & \multirow{2}{*}{\textbf{Dim}} & \multicolumn{3}{c}{\textbf{Accuracy} ($\uparrow$)} \\
    \cmidrule(lr){3-5}
     & & \textbf{MMAU} & \textbf{MMSU} & \textbf{OBQA} \\
    \midrule
    
    \rowcolor{pyellow} 
    \textbf{8 choose 4 (Ours)} & \boldmath$D$ & \textbf{61.50} & \textbf{38.19} & \textbf{53.85} \\
    
    16 choose 4 & $D$   & 60.80 & 34.77 & 48.35 \\
    8 choose 1  & $D$   & 59.89 & 32.95 & 43.95 \\
    4 choose 2  & $D$   & 61.10 & 38.12 & 52.74 \\
    8 choose 4  & $D/2$ & 60.60 & 37.28 & 54.72 \\
    8 choose 4  & $2D$  & 59.69 & 35.94 & 50.54 \\
    
    \bottomrule
    \end{tabular}
\end{table*}

Notably, increasing the total number of experts does not necessarily improve performance: expanding from ``8 choose 4'' to ``16 choose 4'' consistently degrades performance across all benchmarks. 
Regarding routing sparsity, stronger constraints are not always beneficial; activating too few experts, such as ``8 choose 1'', significantly harms audio reasoning ability, while moderate sparsity~(specifically ``4 choose 2'' or ``8 choose 4'') yields stronger results. 
Furthermore, larger expert capacity is not inherently better: increasing expert dimensionality~(``8 choose 4, $2D$'') leads to consistent performance drops, whereas reduced capacity can remain competitive on certain reasoning benchmarks. 
Crucially, under the same training budget, configurations with larger effective capacity through more active experts~(exemplified by ``4 choose 2'' vs.\ ``8 choose 4'') can achieve comparable or improved performance, highlighting the importance of balancing model capacity and routing sparsity.

In conclusion, effective audio reasoning demands a careful balance of expert count, sparsity, and capacity, rather than simply scaling any single dimension in isolation.

\subsubsection{Impact of Expert Balance Loss}
\label{sec:ebl_ablation}

To isolate the impact of the expert balance loss ($\mathcal{L}_{\text{aux}}$) on expert utilization and downstream performance, we evaluate a variant trained without this auxiliary term (w/o EBL). Table~\ref{tab:ablation_ebl} summarizes the results across MMAU, MMSU, and OBQA.

\begin{table}[htbp]
    \centering
    \small
    \setlength{\tabcolsep}{12pt}
    \caption{\textbf{Ablation on Expert Balance Loss.} We report accuracy (\%) on audio reasoning and understanding benchmarks.}
    \label{tab:ablation_ebl}
    
    \begin{tabular}{l ccc}
    \toprule
    \textbf{Method} & \textbf{MMAU} & \textbf{MMSU} & \textbf{OBQA} \\
    \midrule
    
    \rowcolor{pyellow} 
    \textbf{Ours (w/ EBL)} & 61.50 & \textbf{38.19} & \textbf{53.85} \\
    
    w/o EBL                & \textbf{63.01} & 37.37 & 52.31 \\
    \bottomrule
    \end{tabular}
\end{table}

Removing the balance loss yields divergent outcomes across benchmarks. On semantic reasoning tasks, the model incorporating EBL consistently outperforms the ablated counterpart by \textbf{0.82} and \textbf{1.54} points, respectively. This validates that load balancing effectively mitigates expert collapse, ensuring diverse experts are sufficiently trained---a prerequisite for handling complex semantic reasoning grounded in world knowledge.

Conversely, the w/o EBL variant achieves higher accuracy on MMAU. We attribute this behavior to the intrinsic heterogeneity of MMAU, which spans multiple audio modalities (speech, sound, and music) while being dominated by perceptual and low-level acoustic patterns in a large portion of instances. In the absence of explicit load balancing, the router naturally concentrates updates on a subset of ``dominant experts'' that are well aligned with these frequently occurring perceptual and low-level acoustic patterns, resulting in improved performance on perception-oriented audio understanding tasks. However, this concentration of expert utilization reduces the effective diversity of experts participating in optimization, which limits the model’s ability to support systematic generalization and higher-level semantic reasoning. This effect is reflected in the degraded performance observed on MMSU and OBQA, where successful prediction requires grounding audio inputs in world knowledge and performing more complex semantic inference.

In summary, the expert balance loss imposes a trade-off: it slightly constrains perception-oriented performance to significantly enhance semantic reasoning by ensuring balanced expert utilization. We next analyze the routing dynamics to elucidate this mechanism.

\section{Analysis of Expert Specialization and Optimization Dynamics}
\label{sec:analysis}

The MoE-adapter is motivated by the heterogeneous nature of audio data,
where different acoustic patterns and reasoning requirements benefit from specialized representations.
In this section, we analyze the internal behaviors of the MoE-adapter
to explain its performance gains over dense adapters,
as well as the trade-offs observed when introducing expert balance regularization.
Our analysis focuses on two complementary aspects:
(i) how routing-based expert specialization emerges across heterogeneous audio modalities,
and (ii) how such specialization reshapes the optimization dynamics by mitigating gradient conflicts.

\subsection{Effect of Expert Balance on Routing and Specialization}
\label{sec:expert_activation}
\begin{figure}[htbp]
    \centering
    \begin{minipage}{0.48\textwidth}
        \centering
        \includegraphics[width=\linewidth]{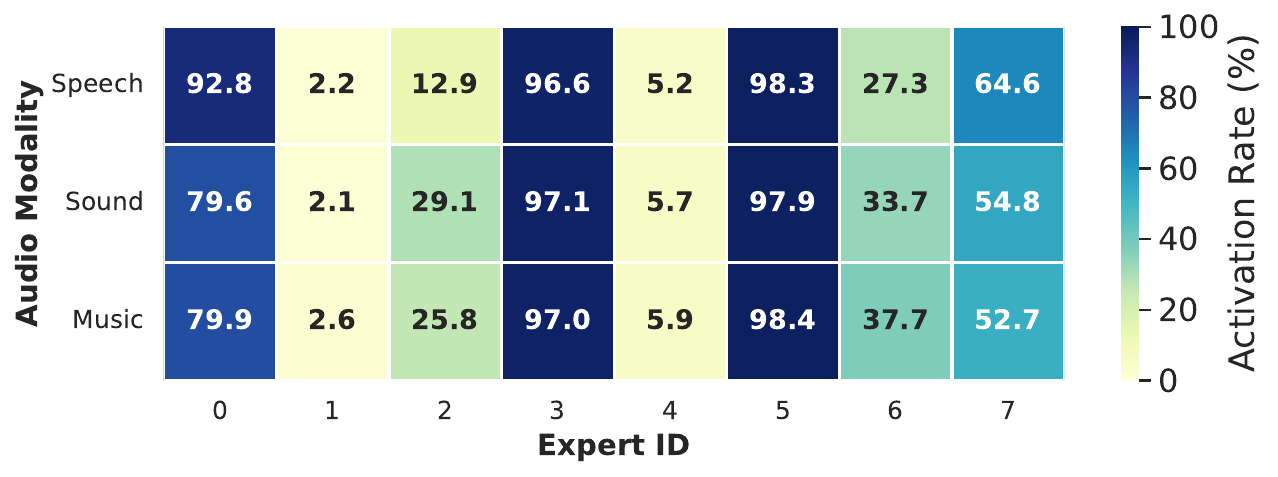}
        \centerline{(a) Without Expert Balance Loss}
    \end{minipage}
    \hfill
    \begin{minipage}{0.48\textwidth}
        \centering
        \includegraphics[width=\linewidth]{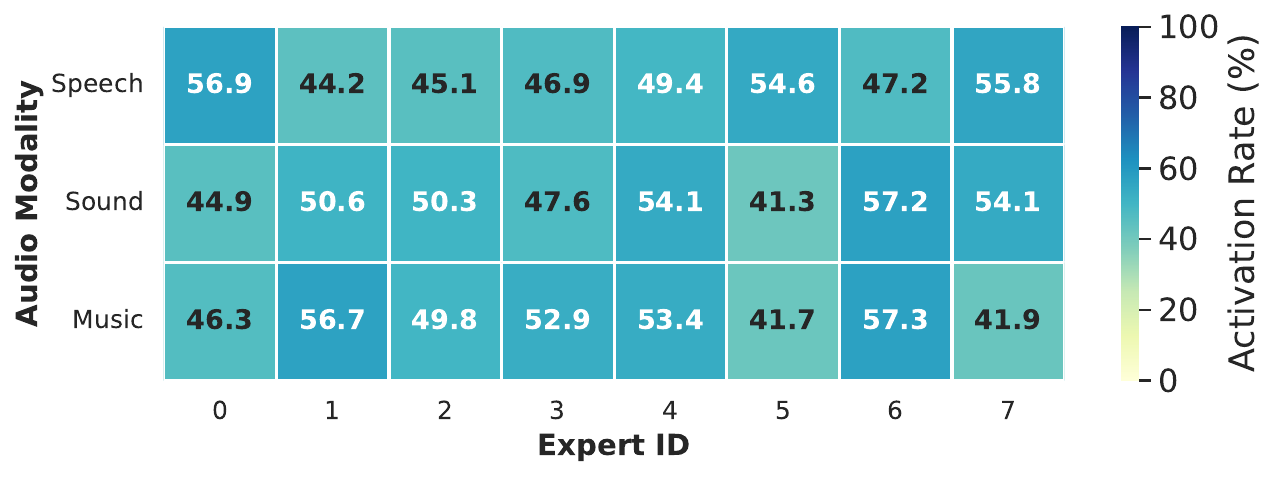}
        \centerline{(b) With Expert Balance Loss (Ours)}
    \end{minipage}

    \caption{\textbf{Expert Activation Heatmaps.} The expert balance loss prevents collapse, fostering balanced utilization while maintaining modality-aware specialization. Notably, Sound serves as an ``Acoustic Bridge'' sharing experts with both Speech and Music.}
    \label{fig:expert_heatmap}
\end{figure}

To understand how the MoE adapter realizes feature disentanglement in practice,
we analyze expert activation patterns on the MMAU,
where samples are categorized into Speech, Sound, and Music.
Figure~\ref{fig:expert_heatmap} visualizes the token-normalized activation rates of routable experts,
aggregated over forward-pass routing decisions.
This comparison directly relates to the ablation results in Section~\ref{sec:ebl_ablation},
where the model without expert balance loss achieves higher performance on the heterogeneous MMAU benchmark.
We observe clear expert specialization across audio modalities.
Certain experts are predominantly activated by a single category,
while others are shared between Sound and either Speech or Music.
Notably, no expert is jointly specialized for both Speech and Music.
This pattern reflects the intrinsic heterogeneity of audio data:
although Sound shares low-level acoustic characteristics with both domains,
Speech and Music differ substantially in temporal structure and semantic organization.
As a result, Sound-centric experts naturally emerge as an acoustic bridge,
enabling controlled parameter sharing without forcing incompatible representations to be jointly optimized.

Importantly, expert balance loss does not suppress specialization,
but rather regularizes its extent.
Compared to the unbalanced setting, it prevents extreme expert dominance
while preserving modality-aware routing behavior.
This regularization introduces a clear trade-off:
while stronger expert dominance in the w/o EBL setting
can benefit broad and heterogeneous benchmarks such as MMAU
by concentrating capacity on a small subset of dominant experts,
it reduces expert diversity and negatively impacts reasoning-intensive tasks.
This emergent expert decomposition provides concrete evidence
that our design motivation—explicitly accounting for audio heterogeneity.
By allowing modality-aware specialization,
the MoE adapter aligns heterogeneous acoustic inputs into a representation space
that is more compatible with the underlying language model.

Beyond routing behavior, such specialization also manifests in the optimization dynamics.
In the following section, we analyze how expert-based routing mitigates gradient conflicts
that commonly arise when learning from heterogeneous audio sources.

\subsection{Gradient Conflicts and Mitigation Mechanisms}
\label{sec:gradient_conflicts}

The adoption of the MoE architecture is primarily motivated by the need to resolve gradient conflicts—a phenomenon frequently observed in dense models when learning heterogeneous tasks. 
In the context of LALMs, the adapter must simultaneously encode divergent modalities: Speech~(semantic-dominant) versus Music and Sound~(paralinguistic-dominant). To rigorously quantify how the proposed MoE Adapter mitigates these optimization bottlenecks compared to a dense FFN baseline, we analyze the training dynamics using two complementary metrics: Gradient Cosine Similarity and the Gradient Influence Score.

\paragraph{Gradient Cosine Similarity.}

\begin{figure}[htbp]
    \centering
    \includegraphics[width=0.7\linewidth]{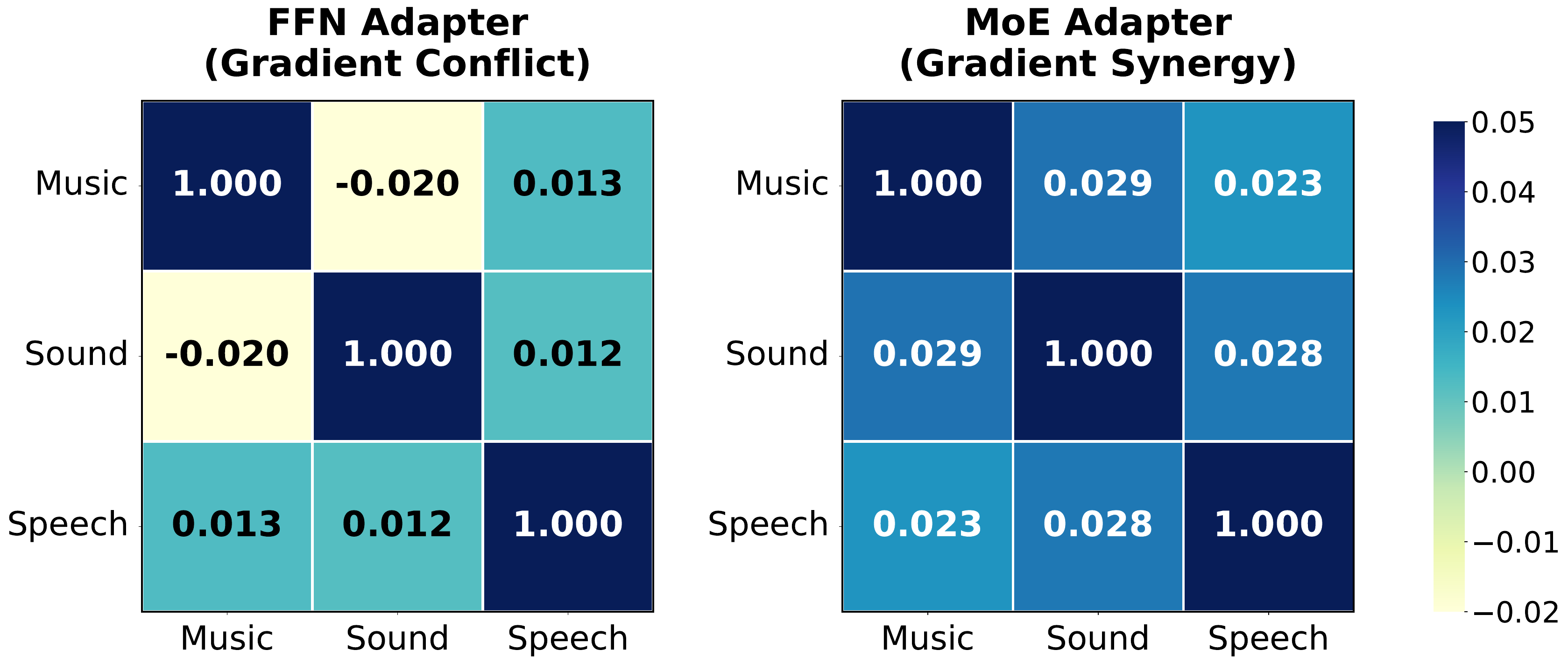}
    
    \caption{\textbf{Gradient Cosine Similarity.} 
    The dense FFN adapter (left) exhibits negative correlations between heterogeneous audio tasks, 
    whereas the MoE adapter (right) demonstrates improved gradient alignment, 
    indicating reduced optimization conflicts.}
    \label{fig:gradient_cosine}
\end{figure}

To investigate optimization interference, we analyze the geometric alignment of gradient vectors across audio categories. 
Specifically, for any two categories $i$ and $j$, we compute the pairwise cosine similarity of their gradients:
\begin{equation}
    \text{Sim}(g_{i}, g_{j}) = \frac{g_{i} \cdot g_{j}}{\|g_{i}\| \|g_{j}\|}
    \label{eq:cosine_sim}
\end{equation}
As illustrated in Figure~\ref{fig:gradient_cosine}~(Left), the monolithic adapter exhibits persistent negative similarities, implying that the optimization direction for one modality is geometrically opposed to another. 
This negative correlation confirms the presence of destructive interference, where minimizing the loss for one task inadvertently hinders the optimization of others.

In contrast, the MoE-Adapter effectively resolves these conflicts, as evidenced by the shift towards positive gradient similarities in Figure~\ref{fig:gradient_cosine}~(Right) (e.g., Music-Speech improves to $+0.023$). 
This emergent orthogonality indicates that the dynamic router successfully decouples the parameter space, directing conflicting gradients to distinct experts. 
This ``subspace specialization'' enables different experts to capture specific acoustic features independently, thereby mitigating the oscillatory dynamics observed in the baseline.

\paragraph{Gradient Influence Analysis.}

\begin{figure}[htbp]
    \centering
    \includegraphics[width=0.7\linewidth]{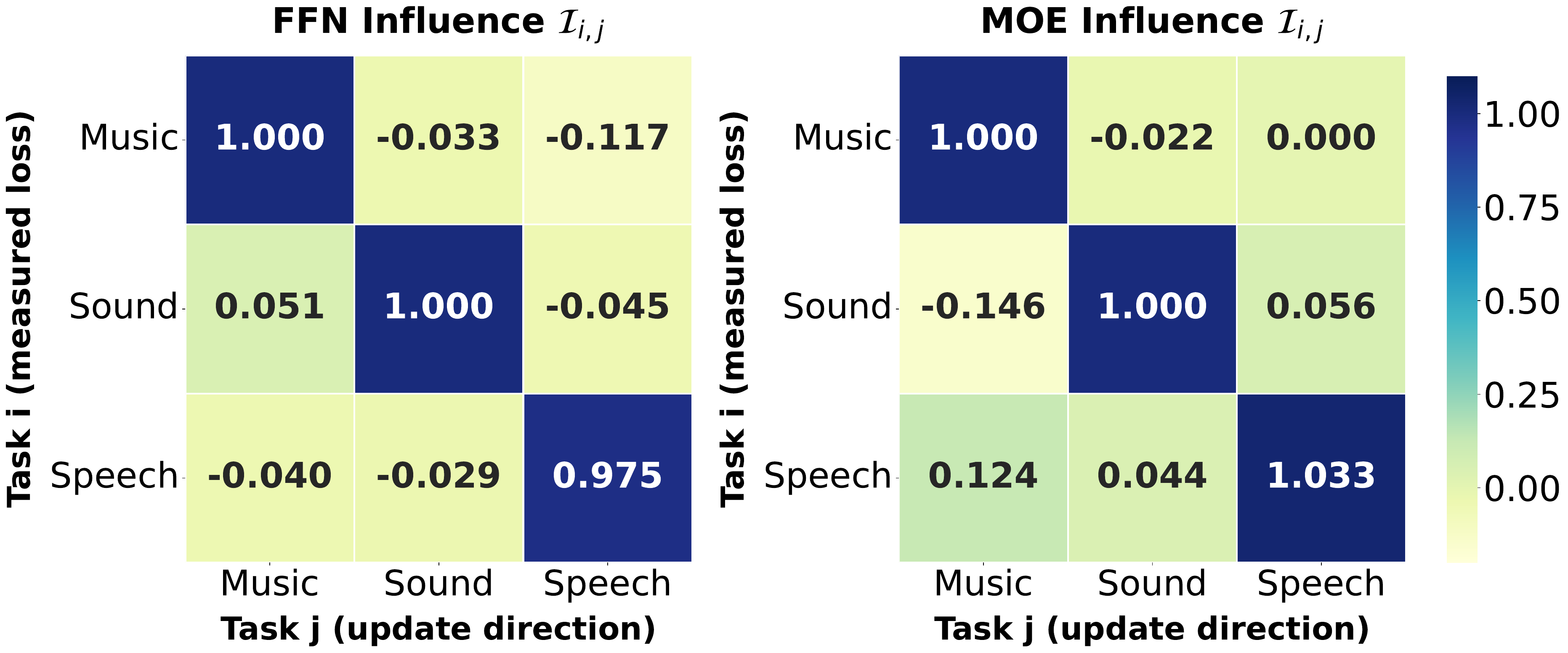}
    
    \caption{\textbf{Gradient Influence Analysis.} Values represent the impact of updating task $j$ (x-axis) on task $i$ (y-axis). Unlike the FFN baseline (left) which suffers from harmful interference (negative values), the MoE adapter (right) promotes constructive transfer (positive values), effectively mitigating cross-task conflicts.}
    \label{fig:gradient_influence}
\end{figure}

To move beyond geometric alignment and measure the causal impact of a gradient update from one task on the loss of another, we employ the Influence Score $\mathcal{I}_{i,j}$. 
Let $\theta_{t}$ denote the current model parameters. 
We simulate a hypothetical update using only the gradient from task $j$, resulting in 
$\theta_{t+1}^{(j)} = \theta_{t} - \lambda \frac{\nabla_{\theta} \mathcal{L}_{j}}{\|\nabla_{\theta} \mathcal{L}_{j}\|}$. 
The resulting reduction in loss for task $i$ is denoted as $\Delta_{j}\mathcal{L}_{i} = \mathcal{L}_{i}(\theta_{t}) - \mathcal{L}_{i}(\theta_{t+1}^{(j)})$. 
The normalized influence score is given by:
\begin{equation}
    \mathcal{I}_{i,j} = \mathbb{E}_{\mathcal{D}}\left[ \frac{\Delta_{j}\mathcal{L}_{i}}{\Delta_{i}\mathcal{L}_{i}} \right]
    \label{eq:influence_score}
\end{equation}
where the expectation is taken over the validation batch. 
Intuitively, a positive score ($\mathcal{I}_{i,j} > 0$) indicates constructive synergy~(updating on $j$ benefits $i$), while a negative score implies destructive interference.

The results, presented in Figure~\ref{fig:gradient_influence}, reveal a sharp contrast. 
The Dense Adapter~(Left) suffers from significant negative interference; notably, updates derived from Speech data actively degrade performance on Music and Sound. 
Conversely, the MoE Adapter~(Right) exhibits predominantly positive influence scores. 
For instance, updates on Speech induce a strong positive transfer to Sound, while interference on Music is substantially mitigated. 
This suggests that the MoE architecture does not merely isolate conflicts; by retaining shared experts for common acoustic features, it fosters a selective positive transfer mechanism, allowing the model to leverage commonalities between Speech and Sound while insulating the distinct attributes of Music.

\section{Conclusion}
We introduce the MoE Adapter to resolve gradient conflicts in LALMs caused by audio heterogeneity. By leveraging dynamic expert specialization, our sparse architecture consistently outperforms dense baselines across diverse benchmarks while maintaining comparable latency. Our analysis confirms that this mechanism transforms destructive gradient interference into constructive updates, underscoring the pivotal role of sparsity as a scalable pathway for future multimodal foundation models.

\section{Limitations}
Although the proposed MoE-Adapter demonstrates superior representational capacity compared to the dense FFN baseline and effectively mitigates gradient conflicts, this work has limitations regarding scalability and generalization. 
First, our evaluations are currently confined to the Qwen3-1.7B backbone; the universality of our method across different LLM families and larger parameter scales (e.g., 70B) remains to be empirically verified. 
Second, we have not explicitly explored the scaling laws of the sparse routing mechanism with respect to training data volume, leaving the performance trajectory of expert specialization under massive-scale training for future investigation.
Finally, we focus on audio understanding and reasoning tasks, and have not yet extended the framework to generative audio applications.

\clearpage
\bibliography{ref}
\bibliographystyle{colm2024_conference}

\clearpage
\appendix
\section{Hyperparameter Configuration}
\label{app:hyperparams}

\begin{table}[h]
    \centering
    \small
    \renewcommand{\arraystretch}{1.3}
    
    \caption{\textbf{Detailed Hyperparameter Configuration.} Both architectures are constrained to the same parameter budget ($\approx$ 94M). Note that $d=2560$ denotes the audio hidden size.}
    \label{tab:hyperparams_robust}
    
    \begin{tabular}{@{} p{0.38\linewidth} p{0.62\linewidth} @{}}
        \toprule
        \textbf{Hyperparameter} & \textbf{Specification} \\
        \midrule
        
        \multicolumn{2}{l}{\textit{\textbf{I. Common Audio Frontend}}} \\
        Speech Encoder Dim & $5,120$ \\
        Audio Hidden Size & $d = 2,560$ \\
        Feature Fusion & $\text{LN} \to \text{Linear}_{5120 \to d} \to \text{SiLU} \to \text{LN}$ \\
        \midrule
        
        \multicolumn{2}{l}{\textit{\textbf{II. Baseline: Dense FFN Adapter}}} \\
        Architecture & Monolithic MLP \\
        Projection Structure & $\text{LN} \to \text{MLP}_{d \to 20480 \to d}$ \\
        Total Parameters & 94.4 M (100\% Active) \\
        \midrule
        
        \multicolumn{2}{l}{\textit{\textbf{III. Ours: MoE Adapter}}} \\
        Architecture & Sparse Mixture-of-Experts \\
        Expert Config & $N=8$ (Total), $k=4$ (Active) \\
        Expert Structure & $\text{LN} \to \text{Experts}_{d \to 1280 \to d} \xrightarrow{\text{Sum}} \text{Out}$ \\
        Gate Network & $\text{Linear}_{d \to 8}$ \\
        Aggregation Block & $\text{LN} \to \text{MLP}_{d \to 10240 \to d}$ \\
        \textbf{Active Parameters} & 70.8 M ($\sim$75\% of Baseline) \\
        \bottomrule
    \end{tabular}
\end{table}

To ensure a rigorous evaluation of the architectural benefits of our proposed method, we conduct a controlled comparison between the Dense FFN Baseline and the MoE-Adapter. We explicitly align the total parameter budget of both models to approximately $94\text{M}$ to isolate the gains attributed to sparse expert specialization. The detailed specifications are summarized in Table~\ref{tab:hyperparams_robust}.

\paragraph{Common Audio Frontend.}
Both models utilize an identical frontend configuration to process acoustic signals. The continuous representations extracted by the speech encoder ($d_{enc}=5,120$) are projected into the model's audio hidden space ($d_{model}=2,560$) via a shared Feature Fusion module. This module consists of a Layer Normalization (LN) layer, a linear projection ($5,120 \to 2,560$), and a SiLU activation function, followed by a final LN layer.

\paragraph{Adapter Architecture Specifics.}
The core divergence lies in the parameter allocation strategy within the projection block:
\begin{itemize}
    \item \textbf{Dense FFN Baseline:} This model employs a monolithic Multi-Layer Perceptron (MLP). To utilize the full parameter budget, we scale the intermediate expansion dimension to $20,480$. Consequently, all $94.4\text{M}$ parameters are active for every input token.
    \item \textbf{MoE-Adapter (Ours):} We decompose the dense projection into $N=8$ specialized experts, each with a smaller intermediate dimension of $1,280$. A learnable gating network routes tokens to the Top-$k=4$ most relevant experts. Distinctively, we introduce a post-routing Aggregation Block---an MLP with an intermediate dimension of $10,240$---to align the aggregated expert outputs with the LLM's embedding space.
\end{itemize}

\paragraph{Computational Efficiency.}
As demonstrated in the statistics section of Table~\ref{tab:hyperparams_robust}, although both architectures maintain a consistent storage footprint ($\sim 94.5\text{M}$), the MoE Adapter significantly reduces computational overhead. During inference, only the routed experts are activated, resulting in an effective parameter count of 70.8 M. This represents a $\sim 25\%$ reduction in active parameters compared to the dense baseline, achieving a superior trade-off between model capacity and inference latency.
\clearpage
\end{document}